%% file: sample-acmsmall.tex
\documentclass[sigconf]{acmart}

\usepackage{multirow}
\usepackage{tcolorbox}
\usepackage{enumitem}

\usepackage{float}
\usepackage{graphicx}
\usepackage{physics}

\newcommand{\verde}[1]{\textcolor{teal}{#1}}
\newcommand{\rojo}[1]{\textcolor{purple}{#1}}
\newcommand{\azul}[1]{\textcolor{blue}{#1}}
\newcommand{\naranja}[1]{\textcolor{orange}{#1}}

\AtBeginDocument{%
  \providecommand\BibTeX{{%
    \normalfont B\kern-0.5em{\scshape i\kern-0.25em b}\kern-0.8em\TeX}}}

\setcopyright{acmlicensed}
\copyrightyear{2024}
\acmYear{2024}
\begin{document}

\title{Quantum software experiments: A reporting and laboratory package structure guidelines}

\author{Enrique Moguel}
\affiliation{%
  \institution{University of Extremadura}  
  \city{Cáceres}
  \country{Spain}
}
\email{enrique@unex.es}
\orcid{0000-0002-4096-1282}

\author{José Antonio Parejo}
\authornotemark[1]
\email{japarejo@us.es}
\author{Antonio Ruiz-Cortés}
\email{aruiz@us.es}
\affiliation{%
  \institution{SCORE Lab, University of Sevilla} 
  \city{Sevilla}
  \country{Spain}}

\author{Jose Garcia-Alonso}
\email{jgaralo@unex.es}
\author{Juan Manuel Murillo}
\email{juanmamu@unex.es}
\affiliation{%
  \institution{University of Extremadura}  
  \city{Cáceres}
  \country{Spain}
}

\renewcommand{\shortauthors}{Moguel and Parejo, et al.}

\begin{abstract}
    \textbf{Background}. In the realm of software engineering, there are widely accepted guidelines for reporting and creating laboratory packages. Unfortunately, the landscape differs considerably in the emerging field of quantum computing. To the best of our knowledge, no standardized guidelines exist for describing experiments or outlining the necessary structures for quantum software laboratory packages.  
    \textbf{Aims}. This paper endeavors to enhance the replicability and verifiability of quantum software experiments. 
    \textbf{Method}. This objective is pursued through the proposition of guidelines for reporting and the delineation of a structure for laboratory packages tailored to quantum computing experiments. Specifically, we advocate for an extension and adaption of established guidelines in experimental software engineering, integrating novel elements to address the specific requirements of quantum software engineering.   
    \textbf{Results}. In validating the utility and effectiveness of the proposed guidelines, we conducted a review encompassing 11 works (5 focusing on reporting guidelines and 6 on laboratory packages). In particular, this review highlighted the absence of standardised guidelines and structure of laboratory packages for quantum software experiments.
    \textbf{Conclusions}. Our assessment revealed gaps in information and opportunities for enhancement within the evaluated papers and laboratory packages. Our proposal contributes to the advancement of quantum software engineering research, taking a fundamental step toward fostering rigorous and reliable scientific research in this emerging paradigm.

\end{abstract}



\received{?? lets see ??}
\received[revised]{??}
\received[accepted]{Who knows..}

\maketitle

\input{intro}
\section{Background}
\label{sec:background}

In this section, we would like to review some related work and introduce some key concepts about quantum computing such as the concepts of qubit, quantum gate, quantum circuit, and quantum algorithm. But on top of the previously defined concept of QSE, we would also like to define the concept of quantum LP, taking advantage of our experience and background in the fields of quantum software engineering and experimental software engineering.

\subsection{Quantum Bit (Qubit)}

A qubit, short for quantum bit, serves as the fundamental unit of quantum information within quantum computing and is distinguished from classical bits by its remarkable properties. While classical bits can exist in either a state of \textit{0} or \textit{1}, qubits can inhabit a superposition of both states concurrently. This superposition grants qubits the ability to represent information not just as a singular value of \textit{0} or \textit{1}, but as a probabilistic combination of both. Mathematically, this superposition is expressed as $\alpha\ket{0}$ + $\beta\ket{1}$, where $\alpha$ and $\beta$ are complex probability amplitudes representing the likelihood of measuring the qubit in the states $\ket{0}$ and $\ket{1}$ respectively. Consequently, a qubit embodies a rich spectrum of potential states, allowing for the encoding of exponentially more information compared to classical bits. Moreover, qubits possess another distinctive property known as entanglement, where the state of one qubit becomes intertwined with the state of another, even when physically separated. 

These properties, superposition, and entanglement, allow us to create highly interconnected quantum systems, which facilitate parallel computing and offer the possibility of solving complex problems that exceed the capabilities of classical computers 
\cite{Aaronson2008}. This is because the amount of information that a qubit system can represent grows exponentially. For example, in a classical computer, you would have to run a function \textit{f} $2^n$ times, once for each input. Whereas in a quantum computer, we can prepare the \textit{n} qubits in a superposition of all $2^n$ possible inputs, and then apply the function \textit{f} to all of them at once. This is called quantum parallelism and can speed up some algorithms exponentially.


\subsection{Quantum Gate}

A quantum gate is a fundamental building block in quantum software, analogous to classical logic gates in classical computing. However, while classical logic gates manipulate bits, quantum gates operate on qubits, the basic units of quantum information. Quantum gates perform operations on qubits to manipulate their quantum states, such as changing their probabilities of being measured in certain states or entangling them with other qubits. Examples of quantum gates include \textit{Hadamard gate} (superposition gate), \textit{Pauli-X gate} (bit-flip gate), or \textit{CNOT gate} (controlled-NOT gate). These gates play a crucial role in executing quantum algorithms by enabling the manipulation and transformation of qubits in a quantum system.

\subsection{Quantum Circuit}

A quantum circuit is a graphical representation of a sequence of quantum operations, analogous to the classical digital circuits of classical computing. In a quantum circuit, quantum gates are arranged sequentially, and qubits flow through them to perform quantum operations. The flow of information and quantum operations within the circuit is represented by lines representing qubits and boxes representing quantum gates. Quantum circuits illustrate the step-by-step execution of a quantum algorithm, showing how qubits are manipulated and transformed throughout the computational process. By designing and implementing quantum circuits, quantum programmers can develop and execute various quantum algorithms to solve quantum-accelerated computational problems.

\begin{figure}[h]
   \includegraphics[width=0.4\textwidth]{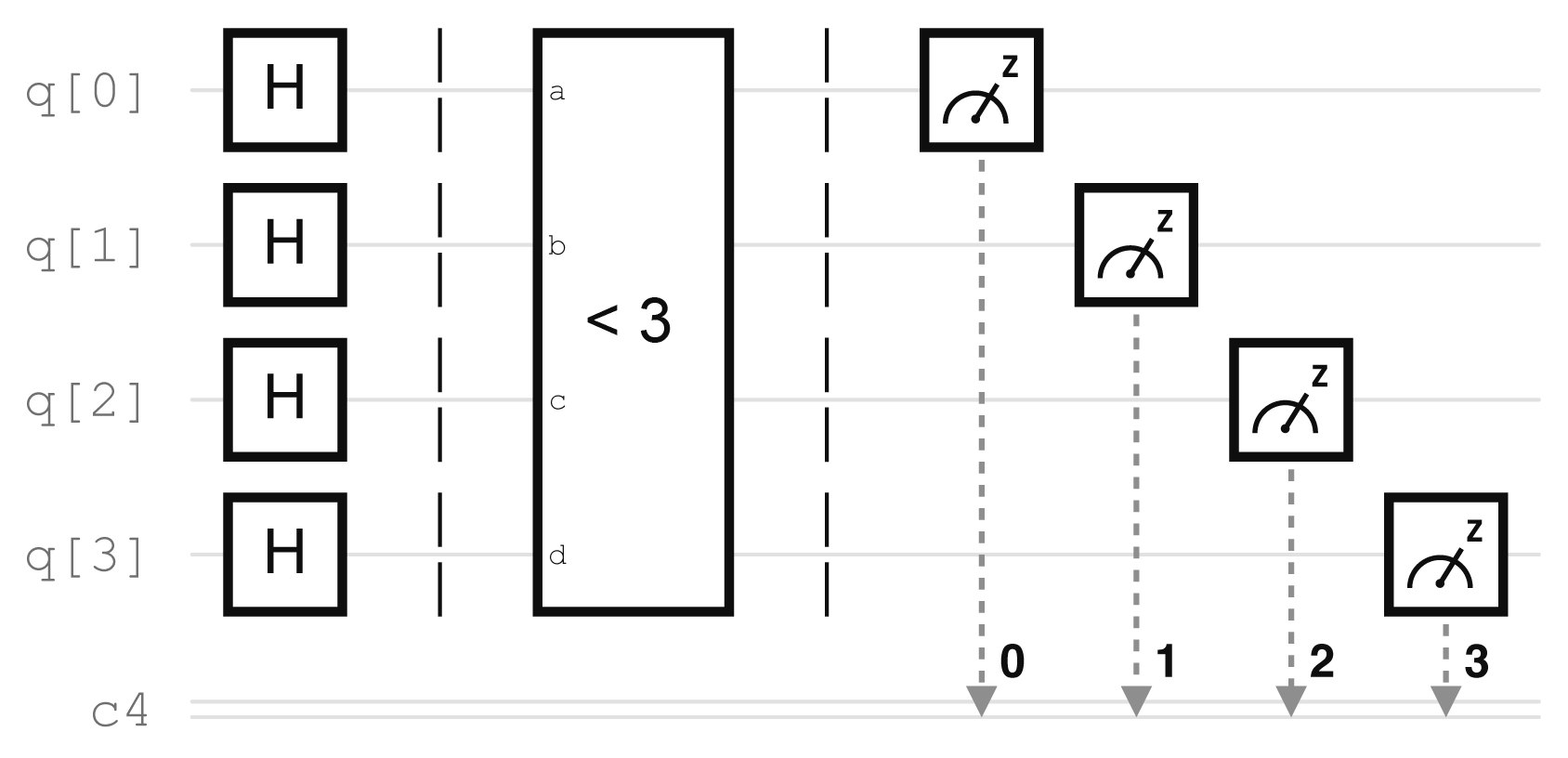}
   \caption{Example of a quantum circuit}
   \Description[Example of a quantum circuit]{Example of a quantum circuit}
   \label{fig:Less3}
\end{figure}

Then, in Figure \ref{fig:Less3}, we present an example of a quantum circuit, in which we can see that we have 4 qubits (from $q_0$ to $q_3$), so we can have results from \textit{0000} to \textit{1111}. Next, we put all these qubits in superposition making use of the \textit{Hadamard quantum gate} (H gate). Subsequently, with the \textit{quantum oracle} (black-box function in which functionality is described) with the function less than the number 3, we mark all those cases where this function is fulfilled. Finally, we collapse the system with the \textit{measurement gate} (Z measurement) and write the results on a classical register (in our case $c_4$). To run this circuit or any other, in a simulator or quantum computer, a number of shots (single runs) must be launched to take advantage of the potential of this computing paradigm and to mitigate the noise that quantum computers can cause. When running this circuit in a simulator or quantum computer, the result that will return us is \textit{0000} $\approx$ 33\%, \textit{0001} $\approx$ 33\%, \textit{0010} $\approx$ 33\%, and the rest of the results $\approx$ 0\%.

\subsection{Quantum Algorithm}

A quantum algorithm is an algorithm performed for a quantum computer in which the states represent a superposition of physical states, the algorithm performs linear and unitary transformations of the states that allow constructive or destructive interference between the physical states, making use of the quantum operations/gates seen in the previous subsection. Some of the most renowned quantum algorithms include \textit{Shor's Algorithm} \cite{Shor1994}, famous for efficiently factoring large integers; \textit{Grover's Algorithm} \cite{grover1996}, known for its quadratic speedup in unsorted database search; or the \textit{Quantum Approximate Optimization Algorithm (QAOA)}, designed to tackle combinatorial optimization problems by leveraging superposition and entanglement. 

Inspired by the analysis of the similarities and differences between classical and quantum programming, there are already proposals that adopt this promising approach to change the level of abstraction of quantum programming languages. This approach consists of treating quantum registers as encodings of data types and basic operations on these data types, such as the oracle (function) defined in Figure \ref{fig:Less3}.

This approach has been explored by considering quantum registers similar to those known in the classical world, such as integer encodings. To subsequently compose simple operations on them, such as the function "\textit{less\_than}", "\textit{is\_even}", "\textit{is\_prime}", etc.

An example, to facilitate the understanding of what a quantum circuit is, can be the greater than and less than operation of a certain number. These operations take as input a quantum state in which the qubits encode natural numbers (including 0). Then, the oracles give a $\pi$-phase to those numbers larger or smaller, respectively, than a given one. With this solution, in a single run, we would have all the possibilities of a number that is greater than and less than one given.

For this example algorithm \cite{Sanchez-Rivero2023} (\textit{less\_than} algorithm, Figure \ref{fig:Less3}), a quantum state in which qubits encode natural numbers is taken as input. Then, the oracle gives a $\pi$-phase to those numbers less than a given one. In addition, in previous work \cite{Sanchez-Rivero2023b}, a classical algorithm was provided that automatically generates the quantum circuit corresponding to the oracles for any number of qubits and input.

\subsection{Quantum Laboratory Packages}

Laboratory Packages have a fundamental role in experimental research, serving not only as a conduit for knowledge transfer and the facilitation of knowledge replication but also as a catalyst for adapting experiments to diverse contexts. Additionally, they offer the prospect of aggregating results, aligning with the inherent evolution of research. Through comprehensive documentation and inclusion of essential artifacts, these packages not only enable the seamless dissemination of knowledge but also empower researchers to adapt methodologies to varied settings and consolidate findings as the research landscape progresses.

Therefore, LPs should not be considered as static elements that remain unchanged after being written. On the contrary, they should be dynamic containers that reflect the evolution of an experimental investigation. In this evolutionary dynamic of the experiment, knowledge and evidence about the subject under investigation are also captured.

Undoubtedly, this tool would facilitate the transfer and evolution of quantum computing research. LPs are instruments that have been used in other scientific disciplines and have already been applied to Software Engineering experiments. However, there is no evidence or indication of similar work in the field of Quantum Software Engineering.



Although for decades there have been widely accepted guidelines on how to perform experiments in Software Engineering \cite{wohlin2012experimentation,juristo2013basics}, proposals for reporting experiment results \cite{jedlitschka2008reporting,kitchenham2008evaluating,kitchenham2007guidelines} which are actively maintained and improved \cite{kitchenham2023guidelines}, and proposals for about the standardization and generation of LPs \cite{Guevara2021,Solari2018}, these guidelines do not cover instructions for performing replication and packaging of experimental materials taking into account the peculiarities of Quantum Computing, such as: \textit{uncertainty} (Heisenberg's uncertainty principle), before measurement, qubits are in superposition of possible results, and when measuring qubits their state collapses and one of the possible results is obtained; \textit{quantum parallelism}, because thanks to superposition, quantum algorithms can perform multiple operations in parallel; \textit{decoherence}, qubits are susceptible to errors due to the loss of quantum coherence and the difficulty of preserving quantum information during computation; \textit{quantum interference}, quantum states can interfere with each other in a constructive or destructive way, which makes it possible to improve or decrease the probability of certain results; etc.


Therefore, we believe 
it is necessary to have a universal structure for the documentation of experiments in Quantum Software Engineering. 
However, to the reach of our knowledge no specific guidelines has been proposed for the reporting of QSEs or the structure of Quantum LPs.



\section{Reporting guidelines for QSEs}
\label{Reporting}

Guidelines for the reporting of software engineering experiments, in general, have been proposed \cite{jedlitschka2008reporting,kitchenham2008evaluating} decades ago. Those guidelines provide comprehensive advice on how to report experiments in software engineering, emphasizing the importance of clear, detailed reporting to enhance the understanding, reproducibility, and usefulness of empirical studies. However, those guidelines do not take into account the current state of quantum computing, with a plethora of different providers, quantum computers that do not support all types of gates that can be found in the circuits, and wide use of simulations due to the scarcity, limited availability, and cost of quantum computing hardware. Moreover, the nature of this kind of experiment, where inherent nuisances appear due to the noise generated by the quantum computing hardware itself requires meticulous conduction of the experiments with special care on the reporting of both experimental conditions and results.   


To streamline the application of the tailored guidelines, we will adopt the structure outlined in \cite{jedlitschka2008reporting}, including its delineated sections. For each section, we will specify the additional elements recommended for detailed reporting or the required modifications for the reporting of the items already present in prior guidelines to suit the unique demands of quantum software engineering experiments. \\

\paragraph{Context.}  This section should provide the information necessary to understand whether the research relates to a specific situation or environment. Practitioners need this information to assess if quantum software or algorithms under study would be applicable in their own organizations. 
Consequently, two critical aspects must be detailed when reporting the context of QSEs. Firstly, the \emph{requirements} imposed by the quantum software on the simulator or quantum hardware should be explicitly defined. This includes, for instance, the minimum number of qbits, the specific types of gates used, and the measurements of quantum state of qbits. Secondly, the \emph{limitations} of the computing platform employed for the experiment. This covers any constraints specific to the provider or the computing setup, such as proprietary or unique constructs used that have been applied due to the limitations or characteristics of the quantum computing platform used. For instance, in some cases, transformations to the quantum software --circuits-- must be performed due to the limitations or characteristics of the actual quantum computing hardware used. This is the case of the triple CNOT gate that IBM Quantum\footnote{\url{https://www.ibm.com/quantum}} supports, but AWS Braket\footnote{\url{https://aws.amazon.com/braket}} do not support, which requires decomposing the gate into a set of equivalent simple CNOT gates.\\

\paragraph{Experimental planning.}  This section is critical for outlining the methodology of an experiment, serving as the ``recipe'' for the experiment. Apart from describing the usual details such as the materials and equipment used, for which special care to detail should be taken in the case of QSEs, researchers must pay special attention to the exhaustive description of the \emph{conduction procedure}. For instance, researchers should state clearly if their quantum software or algorithm has any kind of bootstrapping, that should be executed prior to (or after) the experiment or for each experimental replication. Additionally, if the quantum computing provider or execution platform imposes limitations on the scheduling or execution of the quantum software, those should be clearly and explicitly stated, in order to ease the replication of the experiment. For example, the IBM Quantum platform restricts the execution of a maximum of 3 tasks in parallel (in the execution queue) and does not allow launching tasks with more than 20,000 shots or less than 100 shots. In some cases, integrating related sections or detailed materials in appendices, online resources, and LPs is suggested to accommodate length restrictions and enhance clarity.\\
\paragraph{Experimental design.} Apart from the key elements to be described in the experimental design according to \cite{jedlitschka2008reporting}, it's crucial to include specific details tailored to the complexities and unique aspects of quantum computing.  In QSEs, it is crucial to clearly report and justify the \emph{number of shots}, which refers to the number of times a quantum algorithm is executed under each experimental condition and with the same experimental subject. It is important to distinguish between experimental runs or conditions and shots. Shots are repetitions of the same experimental run, using the same algorithm, parameters (such as initial states), and experimental conditions. Due to the probabilistic nature of quantum computing systems, conducting multiple shots is essential for accumulating sufficient data for the statistical analysis of the quantum software performance and behavior. This practice ensures the reliability and validity of the experimental findings by accounting for quantum variability. \\
\paragraph{Execution.} Researchers must detail the execution platform, whether it involves actual quantum hardware or simulators. This includes specifying the software tools, programming languages (such as Qiskit for IBM, Cirq for Google, or Q\# for Microsoft), and the development environments employed, along with their respective versions. For experiments conducted on quantum computers, it is crucial to report the precision level achieved. In cases where simulators are used, the seeds for random-number generation should be disclosed. Given that these details are technical and a complex experiment may involve multiple values, they can be included in LPs.\\ 
\paragraph{Analysis and reporting.} Given the stochastic nature of quantum computing, an appropriate characterization in terms of statistical distributions over the space of possible outcomes for the results obtained should be performed. This usually involves taking special care in reporting not only central tendency measures but also dispersion measures.\\
\paragraph{Threats to validity}. Apart from the usual threats to validity, QSEs could present specific generalizability threats due to the strong dependence on the computing platform and the diversity of technologies available. In this sense an appropriate description of the requirements and execution constraints of the quantum software is essential, taking into account the immaturity of the field and the high speed at which it is improving.

\begin{table}[!ht]
\centering
\resizebox{0.5\textwidth}{!}{  
    \begin{tabular}{|p{1.8cm}|p{11cm}|}
    \hline
    \textbf{Section} & \textbf{Specific details to report} \\
    \hline
    Context & 
    
         Requirements imposed by the quantum software on the hardware or simulator (e.g., minimum number of qubits, gate types).
         \\ \cline{2-2}
       &  Limitations of the computing platform (e.g., provider-specific constraints, modifications to software due to hardware characteristics).
     \\
    \hline
    Experimental  & 
    
         Any pre-experiment or post-experiment processes required by the quantum software. \\ \cline{2-2}
    Planning & Limitations imposed by the computing provider or execution platform used in the experiment in terms of scheduling or execution.
     \\
    \hline
    Experimental Design & 
    
         Number of shots and its justification.
     \\
    \hline
    Execution & 
    
        Documentation of the quantum computing hardware or simulator, and the software tools, programming languages, and development environments.\\ \cline{2-2}
        &  Specific precision provided by the quantum computing platform at the time of the experiment.
     \\
    \hline
    Analysis and  & 
        Proper characterization of results in terms of statistical distributions \\ \cline{2-2}
    Reporting & Reporting on measures of central tendency, dispersion, and variability. \\
    \hline
    Threats to  &  Generalizability threats specific to QSEs, including the dependence on computing platforms and the diversity of technologies.  \\ \cline{2-2}
    Validity    & Description of requirements and execution constraints due to the field's immaturity and rapid development.
     \\
    \hline
    \end{tabular}
}
\caption{Summary of Reporting Items by Section}
\label{table:reporting_items}
\end{table}

Table \ref{table:reporting_items} provides a summary of the items to be reported specifically in the context of QSEs.

To verify the reporting guidelines we have provided a checklist which can be found in Appendix A at the end of this document.

\section{Laboratory Packages for QSEs}
\label{LaboratoryPackage}

From the instrumental point of view, the LP is a document or set of related documents \cite{Shull2002}. The main objectives of the instrument are: to facilitate the replication of experiments, the recording of results, and their subsequent analysis. The knowledge that can be covered by the LP ranges from the high-level motivations for performing the experiment to the operational materials needed to replicate it. 

According to \textit{Solari} \cite{Solari2018}, the proposed LP, for classical software engineering experiments, must meet a number of conditions to be useful for experiment replication: i) It must be feasible to instantiate for software engineering experiments; ii) It must be flexible enough to be used in different experiments; iii) It must be complete with respect to the experimental research process; iv)It must be effective in obtaining useful replications of experiments.

We understand that these conditions should continue to be met for Quantum Software Engineering. That is why, based on the lessons learned from classical Empirical Software Engineering, we define the structure of an LP containing everything needed to perform an experiment in this area. To achieve this, the LP should not only document the replication instructions and constraints but also include all the concrete materials used during the experiment or how they can be generated. With this, we want to strengthen the concept of the LP for knowledge transfer through self-contained documentation.

The LP proposal tries to group in the same structure all the elements necessary to perform a Quantum Software Engineering experiment replication. This proposal allows documenting, in addition to the experiment itself, the studies performed with it.
\paragraph{Article} This section extends beyond the basic scientific paper, encompassing a range of elements essential for understanding and reproducing the empirical study. It details the experimental design, including methodology, parameters, and variables, offering a clear guide for researchers to follow. It also explores the environmental and situational factors that affected the study, providing a comprehensive view of the research context. Furthermore, the list of authors not only acknowledges contributions but also ensures accountability and transparency by identifying those responsible for the research's conception, execution, and analysis. This extensive documentation within the LP is crucial for spreading knowledge and supporting future work in empirical quantum software engineering. \\

\paragraph{Experimental Material.} This section not only provides detailed instructions on the necessary materials for experimentation but also enriches understanding by including conceptual models and illustrative diagrams that explain the theoretical underpinnings of the experiment. These models visually represent variable interrelationships, enhancing comprehension of the experimental basics. It also lists essential materials such as source code, and specifications for computer or quantum simulators, ensuring study reproducibility. The source code details the algorithmic aspects and computational methods critical to the experiment, while the precise configurations of the computing environment aid in accurately recreating experimental conditions. This thorough documentation, ranging from abstract models to concrete source code, forms a complete and transparent repository in the LP. \\
\paragraph{Datasets.} This section documents the data and preparatory steps leading up to the quantum experiment, detailing the conditions before the actual run. This pre-experimental data is crucial for understanding the initial state and the factors affecting subsequent quantum calculations. It also summarizes the specific manipulations of these data, providing a clear description of the experimental setup and any preprocessing performed. The data from the quantum experiment is key, though its perfect reproduction is challenging due to the variability of quantum systems and their dependence on specific quantum computers or providers. \\
\paragraph{Systems.} This section outlines the technical prerequisites for quantum hardware, including system topology, connection types, qubit count, queuing times, and associated costs. It also details the specific limitations and conditions of the selected quantum computer or provider, such as maximum qubit capacity, availability, and decoherence effects. Additionally, it describes the classical software that manages quantum execution, highlighting its role in interfacing with quantum hardware, and explains the quantum software, providing insights into algorithms and quantum circuits. \\
\paragraph{Scripts.} This section provides a detailed description and analysis of the quantum experiment scripts, explaining the rationale behind key decisions made during their development. It also includes a log file or document created by the authors, which meticulously records the execution of these scripts. This log captures the chronological sequence of the experiment, from the start of the scripts to the end of the computation. By detailing the execution process, nuances, and real-time outcomes, the log file becomes an essential resource. It verifies the experiment’s reliability, guides researchers in understanding the event sequence, and assists in replicating or adapting the quantum experiment. \\
\paragraph{Dictionary.} This section outlines the necessary properties of the LP artifacts. It includes a description and inventory of all artifacts (such as Bitbucket or GitHub), download and usage instructions for the classical and/or quantum artifacts, and the relevant licenses. It also details the development environment and configuration guidelines, potentially incorporating virtual environments like Docker to aid researchers in replicating the experiment. Additionally, it should describe how to verify data access and manipulation during the experiment and include the DOI indicator and licensing or distribution rights of the LP.

Table \ref{table:labpack_items} provides a summary of the elements that a quantum software LP should have to facilitate the replication of experiments, the recording of results, and their subsequent analysis. With respect to the LP proposal followed by Solari \cite{Solari2018}, we propose an extension of up to 13 new fields (marked in blue color in Table \ref{table:labpack_items}) that the LPs of a QSE should have. All these fields have to do with this new computing paradigm, such as data processing to adapt them to quantum environments, conditions and constraints of the quantum environment, description of the quantum software, definition of the development environment, etc.

\begin{table}[!ht]
\resizebox{0.5\textwidth}{!}{  
    \begin{tabular}{|l | p{11cm} |}
    
    \hline
    \textbf{Section} & \textbf{Specific details to LP} \\ \hline
    \multirow{4}{*}{Article} & Scientific article \\ \cline{2-2} 
    & Description of the experimental design \\ \cline{2-2} 
    & Description of the experiment context \\ \cline{2-2} 
    & Authors \\ \hline
    
     & Instructions for the experimental material \\ \cline{2-2} 
    Experimental & Conceptual Models and/or representative diagrams (data representativeness) \\ \cline{2-2} 
     Material & \azul{Materials needed to conduct the experiment (source code, quantum computer configuration or provider, features, etc.)} \\ \hline
    
    \multirow{3}{*}{Datasets} & Pre-experimental data \\ \cline{2-2} 
    & \azul{Data processing (for its representation in Quantum)} \\ \cline{2-2} 
    & \azul{Data resulting from the experiment (non-replicability and machine-dependent)} \\ \hline
    
    \multirow{4}{*}{System} & \azul{Selection and definition of hardware technical requirements (topology, connection, number of qubits, queue, cost, etc.)} \\ \cline{2-2} 
    & \azul{Conditions and restrictions of the quantum computer and/or provider (maximum number of qubits, availability, decoherence, etc.)} \\ \cline{2-2} 
    & Description of Classic Software \\ \cline{2-2} 
    & \azul{Description of Quantum Software (gates/annealing, algorithm, number of shots)} \\ \hline
    
    \multirow{3}{*}{Scripts} & \azul{Description of the followed script (including decisions made)} \\ \cline{2-2} 
    & \azul{Script for analyzing the obtained results} \\ \cline{2-2} 
    & Log file of the obtained results (CSV, images, etc.) \\ \hline
    
    \multirow{12}{*}{Dictionary} & Description of the artifacts \\ \cline{2-2} 
    & Artifacts inventory (Bitbucket, GitHub, etc.)  \\ \cline{2-2} 
    & \azul{Instructions for downloading artifacts (classic and/or quantum)} \\ \cline{2-2} 
    & \azul{Instruction guide for executing artifacts (classic and/or quantum)} \\ \cline{2-2} 
    & Artifact license \\ \cline{2-2} 
    & Virtual environments (Docker, Kubernetes, etc.)  \\ \cline{2-2} 
    & \azul{Definition of the development environment} \\ \cline{2-2} 
    & \azul{Guidelines for configuring and accessing the development environment} \\ \cline{2-2} 
    & \azul{Verification of access and functionality} \\ \cline{2-2} 
    & Decisions made during the experiment \\ \cline{2-2} 
    & DOI \\ \cline{2-2} 
    & License or distribution right of the Laboratory Package \\ \hline
    \end{tabular}
}
\caption{Summary of Laboratory Package Items by Section}
\label{table:labpack_items}
\end{table}

To verify the structure and content of the LP we have provided a checklist which can be found in Appendix B at the end of this document.


\section{Validation}
\label{sec:validation}


Quantum Software Engineering is an emerging field with existing research proposing various techniques, tools, and technologies to advance the discipline. However, not all researchers make their derived artifacts publicly available. To validate the applicability and effectiveness of the proposed guidelines, this study specifically focuses on empirical studies conducted by leading researchers in Quantum Software Engineering, and on studies with open LPs, allowing any scientist, researcher, or developer to replicate the research. We selected the papers that were more empirically complete in terms of diversity of scientific contribution and methodology as described below.

It is worth noting that although this work aims to propose reporting guidelines and essential content for publicly available LPs, not all suggested fields are mandatory, nor do they carry the same importance for replicating experiments.


\subsection{Validation of the  Reporting Guidelines}

In the application of the above-stated reporting guidelines, we have undertaken a rigorous analysis of five papers that represent a broad spectrum of the quantum software domain. This selection was intentional, aimed at encompassing a diverse range of QSEs, using both quantum computers simulators \cite{Mendiluze2024,Shahid2024,Wang2021} and actual quantum computers \cite{alvarado2023,Romero-Alvarez2023}. The evaluation process involved a detailed review of each paper, with a particular focus on how well the experiments described adhered to the established guidelines. 

To synthesize the outcomes of our evaluation, we have compiled a comprehensive review in Table \ref{table:reportingGuidelinesEvaluation}
. This table acts as a visual abstract, illustrating how the evaluated QSEs align with each guideline criterion, with papers listed in columns and criteria in rows. It shows the compliance level for each criteria. Detailed explanations follow for any papers or criteria with ambiguous interpretations, indicated by superscripts in the affected table cells.

\begin{table*}[!htp]\centering
\resizebox{0.85\textwidth}{!}{  
    \begin{tabular}{|l|p{9.5cm}|ccccc|c|}\hline
    \textbf{Area}  &\textbf{Papers} &\textbf{\cite{alvarado2023}} &\textbf{\cite{Mendiluze2024}} &\textbf{\cite{Romero-Alvarez2023}} &\textbf{\cite{Shahid2024}} & \textbf{\cite{Wang2021}} & Total \verde{$\checkmark$} \\ \hline
    
    \multirow{6}{*}{Context} &Min Q-bits &
    \verde{\verde{$\checkmark$}} &\verde{$\checkmark^4$} &\verde{$\checkmark^2$} &\verde{$\checkmark$} &\verde{$\checkmark$} & 5 / 5 \\
    & Gate requirements &\verde{$\checkmark$} &\verde{$\checkmark^4$} &\verde{$\checkmark^2$} &\verde{$\checkmark$} &\verde{$\checkmark$} & 5 / 5 \\
    & Measurements &\verde{$\checkmark$} &\verde{$\checkmark$} &\verde{$\checkmark^2$} &\verde{$\checkmark$} &\rojo{$\times$} & 4 / 5 \\
    & Circuit depth (if appropriate) &\verde{$\checkmark$} &\verde{$\checkmark$} &\verde{$\checkmark^2$} &$-^7$ &\verde{$\checkmark$} & 4 / 5 \\
    & Connectivity between qubits &\verde{$\checkmark$} &\verde{$\checkmark^4$} &\verde{$\checkmark^2$} &$-^7$ &\rojo{$\times$} & 3 / 5 \\
    & Limitations imposed by the execution platform (max execution time, max qbits, etc.) &\verde{$\checkmark$} &\verde{$\checkmark$} &\rojo{$\times$} &\verde{$\checkmark$} &\verde{$\checkmark$}  & 4 / 5 \\ \hline
    
    Experimental  &Bootsrapping needs &$-$ &$-^5$ &\verde{$\checkmark$} &\verde{$\checkmark$} &$-^5$ & 2 / 5 \\
    Planning & Target platform execution limitations &\verde{$\checkmark$} &\verde{$\checkmark$} &\rojo{$\times$} &\verde{$\checkmark$} &\verde{$\checkmark$} & 4 / 5 \\ 
    & Target platform planning limitations &$-$ &$-^5$ &\rojo{$\times$} &$-^5$ &$-^5$ & 0 / 5 \\  \hline
    
    Experimental & Justification for the shots and shots to execute &\verde{$\checkmark$} &\verde{$\checkmark$} &\rojo{$\times$} &\rojo{$\times$} &\rojo{$\times^8$} & 2 / 5 \\
    Design& Techniques used to initialize the quantum state &\verde{$\checkmark$} &$-^5$ &\rojo{$\times$} &\verde{$\checkmark$} &$-^5$ & 2 / 5 \\ \hline
    
    \multirow{7}{*}{Execution}  & Precission &\rojo{$\times$} &$-^5$ &\rojo{$\times$} &$-^5$ &$-^5$ & 0 / 5 \\
    & Calibration data for the qubits used during the experiment &\rojo{$\times$} &$-^5$ &\rojo{$\times$} &$-^5$ &$-^5$ & 0 / 5 \\
    & Explicit classical pre-processing and post-processing steps specification&$\sim^1$ &\verde{$\checkmark^4$} &$\sim^1$ &\verde{$\checkmark$} &\verde{$\checkmark^4$} & 3 / 5 \\
    & Quantum programming language and framework (e.g., Qiskit, Cirq, etc.). &\verde{$\checkmark$} &\verde{$\checkmark$} &$-$ &\verde{$\checkmark$} &\verde{$\checkmark$} & 4 / 5 \\
    & The version of the software framework and any dependencies &\verde{$\checkmark$} &\verde{$\checkmark$} &\verde{$\checkmark$} &\verde{$\checkmark$} &\verde{$\checkmark$} & 5 / 5 \\
    & Random seeds used in case of simulation-based experiments & $-$ &\verde{$\checkmark$} & $-$ &\rojo{$\times$} &\rojo{$\times$} & 1 / 5 \\
    & Measurement settings and readout error mitigation techniques. &$-$ &$-^5$ &$-$ &$-^5$ &$-^5$ & 0 / 5 \\ \hline
    
    \multirow{2}{*}{Analysis} & Statistical modelling of the distributions of outputs &\rojo{$\times^3$} &\verde{$\checkmark$} &$-$ &\rojo{$\times$} &\rojo{$\times$} & 1 / 5 \\
    & Quantum error correction or mitigation techniques employed &\verde{$\checkmark$} &$-^5$ &$-$ &$-^5$ &$-^5$ & 1 / 5 \\ \hline
    
    \multirow{2}{*}{Treats to}  &Analysis of how the results might vary with different quantum hardware &$-$ &$-^5$ &\rojo{$\times$} &$-^5$ &$-^5$ & 0 / 5 \\
    Validity& Description for how the experiment might be modified or extended to other quantum computing platforms &\verde{$\checkmark$} &\verde{$\checkmark$} &\verde{$\checkmark$} &\rojo{$\times$} &\rojo{$\times$} & 3 / 5 \\
    \hline

    Total \verde{$\checkmark$} & & 13 / 22 & 14 / 22 & 8 / 22 & 10 / 22 & 8 / 22 & \\ \hline
    \end{tabular}
}
\caption{Reporting guidelines evaluation}
\label{table:reportingGuidelinesEvaluation}
\end{table*}

\begin{enumerate}
    \item The pre-processing and post-processing steps of the algorithms applied are derived from the original papers where they were proposed (as stated, but not specified in the paper)
    \item The proposal introduces a continuous deployment strategy, it does not impose limitations or requirements on quantum computing hardware (stated explicitly in the paper).
    \item No information about discrepancies inter-/intra-provider results is provided. Maybe the error is zero due to solution encoding/decoding, but it is not mentioned in the paper.
    \item They rely on other papers (providing appropriate citations) for the specification of each algorithm, thus we consider it accomplished
    \item They use a simulator for all the experimentation
    \item In this paper most parameters depend on the specific characteristics of the SAT formula to be solved, however, they provide either formulas or statistical approximations for the calculation of the corresponding parameters
    \item We assume it is not necessary, using a well-documented and widely applied algorithms.  
    \item Authors apply a k-wise experimental design with 500 executions, but there is no discussion on how many executions should be performed due to simulation uncertainty or shots.
\end{enumerate}

The results in Table \ref{table:reportingGuidelinesEvaluation} show potential for improvement. The table indicates that 48.2\% of criteria are met (\verde{$\checkmark$}), 30.9\% are not applicable ($-$), 19.1\% are unmet (\rojo{$\times$}), and 1.8\% are partially met ($\sim$). The criteria are best met in the area of Context, while they are less satisfactorily met in Experimental Design, Analysis, and Threats to Validity. Specific criteria that are well met include the version of the software framework and its dependencies and essential context parameters such as the minimum number of qbits and gate requirements. The least satisfactorily met criteria involve the specification of precision and calibration during execution, as well as generalizability threats due to specific limitations of the quantum hardware. The paper that most closely adheres to the reporting guidelines is \cite{Mendiluze2024}, with 63.6\% of the criteria fully met. Additionally, the other 37.4\% of the criteria were deemed not applicable, as the study was conducted using quantum computing simulators. The papers with the most room for improvement in adhering to the reporting guidelines are \cite{Romero-Alvarez2023, Wang2021} with a  36.3\% of criteria fully met. Additionally, \cite{Romero-Alvarez2023} meets the criteria for specifying classical pre-processing and post-processing only partially, directing readers to referenced papers for detailed information, but not specifying the quantum software algorithms with such steps explicitly.

\subsection{Validation of the Laboratory Package}

We have performed a rigorous analysis of LPs representing a diverse spectrum of the quantum software engineering domain. This selection is based on different works of experts in the field of study with the objective of covering a diverse range of QSEs as variable as possible. The evaluation process consisted of the review of LPs and the derived scientific papers, following the LP structure proposed in this work.

The evaluation has been done with 6 different LPs, 3 of these papers \cite{alvarado2023,Mendiluze2024,Romero-Alvarez2023} have also been evaluated in the previous section for proposal information guidelines. More specifically, the LP \cite{Romero-AlvarezZenodo2023} makes reference to the work of \textit{Romero-Alvarez et al.} \cite{Romero-Alvarez2023}; the LP \cite{Alvarado-ValienteZenodo2022} makes reference to the work of \textit{Alvarado-Valiente et al.} \cite{alvarado2023}; and the LP \cite{EnautMendiGithub2023} makes reference to the work of \textit{Mendiluze et al.} \cite{Mendiluze2024}.

To synthesize the results of this evaluation, we have summarized the review in Table \ref{table:LPs_Evaluation}, which presents the results obtained in a clear manner. This table serves as a visual summary of the findings obtained, thus facilitating the understanding of the results obtained from our evaluation and of the status of the LPs for experiments performed by experts in the field of quantum software engineering. For each work, the table indicates its degree of compliance following the package structure that we propose in this work.

\begin{table*}[!htp]\centering
\resizebox{0.85\textwidth}{!}{  
    \begin{tabular}{|l|p{6cm}|cccccc|c|}\hline
    \textbf{Section} & \textbf{Specific details} & \textbf{\cite{Sanchez-RiveroGithub2023}} & \textbf{\cite{QuAntiLGithub2022}} & \textbf{\cite{cda-tumGithub2023}} & \textbf{\cite{Romero-AlvarezZenodo2023}} & \textbf{\cite{Alvarado-ValienteZenodo2022}} & \textbf{\cite{EnautMendiGithub2023}} & Total \verde{$\checkmark$} \\ \hline
    
        Article & Scientific article & \rojo{$\times$} & \verde{$\checkmark$} & \verde{$\checkmark$} & \verde{$\checkmark$} & \verde{$\checkmark$} & \verde{$\checkmark$} & 5 / 6 \\
        & Description of the experimental design & \rojo{$\times$} & \rojo{$\times$} & $-^1$ & $-^1$ & $-^1$ & $-^1$ & 0 / 6 \\
        & Description of the experiment context & \rojo{$\times$} & \verde{$\checkmark$} & \verde{$\checkmark$} & \verde{$\checkmark$} & \verde{$\checkmark$} & \verde{$\checkmark$} & 5 / 6 \\
        & Authors & \rojo{$\times$} & $-^1$ & $-^1$ & \verde{$\checkmark$} & \verde{$\checkmark$} & $-^1$ & 2 / 6 \\
        \hline
    
        Experimental & Instructions for the experimental material & \rojo{$\times$} & \verde{$\checkmark$} & \verde{$\checkmark$} & \verde{$\checkmark$} & \verde{$\checkmark$} & \verde{$\checkmark$} & 5 / 6 \\
        Material & Conceptual Models and/or diagrams & \rojo{$\times$} & \verde{$\checkmark$} & \verde{$\checkmark$} & $-^1$ &\rojo{$\times$} & $-^1$ & 2 / 6 \\
        & Materials to conduct the experiment & \rojo{$\times$} & \verde{$\checkmark$} & \verde{$\checkmark$} & \verde{$\checkmark$} & \verde{$\checkmark$} & \verde{$\checkmark$} & 5 / 6 \\
        \hline
    
        Datasets & Pre-experimental data & \verde{$\checkmark$} & \verde{$\checkmark$} & \verde{$\checkmark$} & \verde{$\checkmark$} & \verde{$\checkmark$} & \verde{$\checkmark$} & 6 / 6 \\
        & Data processing & \rojo{$\times$} & \verde{$\checkmark$} & \rojo{$\times$} & \rojo{$\times$} & \verde{$\checkmark$} & $-^2$ & 2 / 6 \\
        & Data resulting from the experiment & \rojo{$\times$} & \verde{$\checkmark$} & \rojo{$\times$} & \verde{$\checkmark$} & \verde{$\checkmark$} & \verde{$\checkmark$} & 4 / 6 \\
        \hline
    
        Systems & Hardware technical requirements & \rojo{$\times$} & \rojo{$\times$} & \verde{$\checkmark$} & \rojo{$\times$} & $-^3$ & $-^3$ & 1 / 6 \\
        &  Restrictions of the quantum computer & \rojo{$\times$} & \rojo{$\times$} & \rojo{$\times$} & \rojo{$\times$} & \rojo{$\times$} & \verde{$\checkmark$} & 1 / 6 \\
        & Description of Classic Software & \rojo{$\times$} & \verde{$\checkmark$} & \verde{$\checkmark$} & \rojo{$\times$} & $-^3$ & $-^3$ & 2 / 6 \\
        & Description of Quantum Software & \rojo{$\times$} & \rojo{$\times$} & \rojo{$\times$} & $-^3$ & $-^3$ & $-^3$  & 0 / 6 \\
        \hline
    
        Scripts & Description of the followed script & \rojo{$\times$} & \verde{$\checkmark$} & \verde{$\checkmark$} & \rojo{$\times$} & \rojo{$\times$} & \rojo{$\times$} & 2 / 6 \\
        & Script for analyzing the obtained results & \rojo{$\times$} & \rojo{$\times$} & \verde{$\checkmark$} & \rojo{$\times$} & \rojo{$\times$} & $-^2$ & 1 / 6 \\    
        & Log file of the obtained results & \rojo{$\times$} & \verde{$\checkmark$} & \verde{$\checkmark$} & \rojo{$\times$} & \verde{$\checkmark$} & \verde{$\checkmark$} & 4 / 6 \\
        \hline
    
        Dictionary & Description of the artifacts & \rojo{$\times$} & \rojo{$\times$} & \verde{$\checkmark$} & $-^4$ & \verde{$\checkmark$} & \verde{$\checkmark$} & 3 / 6 \\
        & Artifacts inventory & \verde{$\checkmark$} & \verde{$\checkmark$} & \verde{$\checkmark$} & \verde{$\checkmark$} & \verde{$\checkmark$} & \verde{$\checkmark$} & 6 / 6 \\    
        & Instructions for downloading artifacts & \rojo{$\times$} & \verde{$\checkmark$} & \verde{$\checkmark$} & \verde{$\checkmark$} & \verde{$\checkmark$} & \verde{$\checkmark$} & 5 / 6 \\
        & Instruction guide for executing artifacts & \rojo{$\times$} & \verde{$\checkmark$} & \verde{$\checkmark$} & \verde{$\checkmark$} & \verde{$\checkmark$} & \verde{$\checkmark$} & 5 / 6 \\
        & Artifact license & \verde{$\checkmark^5$} & \verde{$\checkmark^5$} & \verde{$\checkmark^5$} & \verde{$\checkmark^5$} & \rojo{$\times$} & \rojo{$\times$} & 4 / 6 \\
        & Virtual environments & \rojo{$\times$} & \rojo{$\times$} & \rojo{$\times$} & \verde{$\checkmark$} & \rojo{$\times$} & \rojo{$\times$} & 1 / 6 \\
        & Development environment definition & \rojo{$\times$} & \rojo{$\times$} & \rojo{$\times$} & $-^4$ & \verde{$\checkmark$} & $-^4$ & 1 / 6 \\
        & Guidelines for development environment & \rojo{$\times$} & \rojo{$\times$} & \rojo{$\times$} & \verde{$\checkmark$} & \verde{$\checkmark$} & \verde{$\checkmark$} & 3 / 6 \\
        & Verification of access and functionality & \rojo{$\times$} & \verde{$\checkmark^6$} & \rojo{$\times$} & \rojo{$\times$} & \rojo{$\times$} & \rojo{$\times$} & 1 / 6 \\
        & Decisions made during the experiment & \rojo{$\times$} & \verde{$\checkmark^6$} & \rojo{$\times$} & \rojo{$\times$} & \rojo{$\times$} & \rojo{$\times$} & 1 / 6 \\
        & DOI & \rojo{$\times$} & \rojo{$\times$} & \rojo{$\times$} & \verde{$\checkmark$} & \verde{$\checkmark$} & \rojo{$\times$} & 2 / 6 \\
        & License of the Laboratory Package & \verde{$\checkmark^5$} & \verde{$\checkmark^5$} & \verde{$\checkmark^5$} & \verde{$\checkmark^5$} & \verde{$\checkmark^5$} & \rojo{$\times$} & 5 / 6 \\
        \hline

        Total \verde{$\checkmark$} & & 4 / 29 & 18 / 29 & 17 / 29 & 15 / 29 & 17 / 29 & 13 / 29 & \\ \hline
        
    \end{tabular}
}
\caption{Laboratory Package structure proposal evaluation}
\label{table:LPs_Evaluation}
\end{table*}

From this evaluation, we have extracted a few details that we believe to be of relevance:

\begin{enumerate}
    \item Some information (such as the description of the experimental design, some conceptual models, or the authors' information) is not detailed in the LPs, but this information can be extracted from the derived papers. 
    
    \item Some LPs do not detail information about data processing or other guides on how the results obtained were arrived at, although the authors provide a step-by-step execution guide using \textit{Python Notebooks}.
    
    \item Some of the analyzed works do not detail all the technical information about the execution environment they have used for their experiments, although they show some information such as some descriptions of the classical software used, some descriptions of the quantum software or hardware used or even the partial definition of the different artifacts that compose the LP.
    
    \item Partial information on the definition of the development environment is provided.
    
    \item Not all works offer their LPs or some specific artifacts under a license. The majority of the authors rely on free licenses such as GPL 3.0, Apache 2.0, or MIT licenses.
    
    \item Only work derived from the QuAntiL project \cite{QuAntiLGithub2022} provides verification of access and correct operation of the LP, in addition to the decisions made during the experiment.
\end{enumerate}

The results in Table \ref{table:LPs_Evaluation} show potential for improvement. The table indicates that 47.7\% of criteria are met (\verde{$\checkmark$}), 12.6\% are partially met ($-$), and 39.7\% are unmet (\rojo{$\times$}). The criteria are met to a greater extent in the areas of Article and Experimental Material, while they are met less satisfactorily in the areas of Datasets, Systems, Scripts, and Dictionary. 

On the other hand, the specific criteria that all works meet (6/6) are those of providing the pre-experimental data used, as well as making available (in repositories such as Github) the artifacts developed for the experiment. The specific criteria that are less met (0 or 1/6) are those of defining the conditions and/or restrictions of the quantum computer and/or the provider used for the experiment (e.g. defining the maximum number of qubits, the availability of the computer/provider, the decoherence, etc.); providing a guide for the verification of access and correct functioning of the LP; or explicitly indicating the decisions taken during the experiment. In addition, of the evaluated works, the work of \textit{QuAntiL project} \cite{QuAntiLGithub2022} is the one that has fulfilled the most fields, being 18 out of a total of 29 fields; while the work of \textit{Range-integers-oracle lab-pack} \cite{Sanchez-RiveroGithub2023} is the one that fulfills the least number of fields with 4 out of 29.


\section{Conclusion}
\label{sec:conclusion}


This paper makes a significant contribution to the field of quantum computing research by introducing a new set of guidelines for reporting QSEs and proposing a structured for quantum laboratory packages. These developments are designed to meet the unique challenges and requirements of this thriving field, aiming to enhance the rigor and reproducibility of research within quantum computing. Additionally, a thorough validation process was conducted, involving 11 quantum software experiments (five focused on the reporting guidelines and six on the structure and contents of the laboratory packages). This process has successfully demonstrated the applicability and effectiveness of the proposed guidelines.

The review of existing work not only identified points for improvement in reporting practices but also highlighted the need for standardized guidelines and structured LPs in quantum software engineering. The proposed framework not only improves the replicability and verifiability of experiments but also fosters a culture of rigorous and reliable scientific research within the quantum computing community.


In future work, the authors aim to expand this evaluation to further substantiate and validate these contributions. This future work will help to map out the current landscape of replicability in quantum software engineering, aiming to foster a more robust and transparent research environment in this promising and rapidly evolving field.

In conclusion, this work constitutes a valuable resource for researchers, scientists, and developers engaged in quantum software engineering. By promoting transparency, consistency, and best practices in reporting QSEs and publishing appropriate LPs, the guidelines described in this paper contribute significantly to the advancement of quantum computing research.

\section*{Acknowledgments}

This work has been partially funded by the European Union “Next GenerationEU/PRTR”, by the Ministry of Science, Innovation and Universities (projectsPID2021- 126227NB-C21, PID2021-126227NB-C22, TED2021-131023B-C21, TED2021-131023B-C22, PID 2021 - 1240454OB-C31, TED2021-130913B-I00, and PDC2022-133465-I00). It is also supported by QSERV: Quantum Service Engineering: Development Quality, Testing and Security of Quantum Microservices project funded by the Spanish Ministry of Science and Innovation and ERDF; by the Regional Ministry of Economy, Science and Digital Agenda of the Regional Government of Extremadura (GR21133); and by European Union under the Agreement — 101083667 of the Project “TECH4E -Tech4effiencyEDlH” regarding the Call: DIGITAL-2021-EDlH-01 supported by the European Commission through the Digital Europe Program. It is also supported by the QSALUD project (EXP00135977 / MIG-20201059) in the lines of action of the Center for the Development of Industrial Technology (CDTI).

\bibliographystyle{ACM-Reference-Format}
\bibliography{sample-base}

\onecolumn

\appendix


\section*{Appendix A: A checklist for the reporting of quantum software experiments}
\label{Appendix_A}

\begin{table*}[!ht]
\centering
\resizebox{0.6\textwidth}{!}{
\begin{tabular}{|p{2cm}|p{10.5cm}|c|}
\hline
\multirow{6}{*}{Context} &Min Q-bits &
$\square$  \\
& Gate requirements &$\square$ \\
& Circuit depth (if appropriate) &$\square$ \\
& Connectivity between qubits &$\square$ \\
& Limitations imposed by the execution platform (max execution time, max qbits, etc.) &$\square$ \\ \hline
Experimental  &Bootsrapping needs &$\square$ \\
Planning & Target platform execution limitations &$\square$ \\ 
& Target platform planning limitations &$\square$ \\  \hline
Experimental & Justification for the shots and shots to execute &$\square$ \\
Design& Techniques used to initialize the quantum state &$\square$ \\ \hline
\multirow{7}{*}{Execution}  &Precission &$\square$ \\
& Calibration data for the qubits used during the experiment &$\square$ \\
& Classical pre-processing and post-processing steps &$\square$ \\
& Quantum programming language and framework (e.g., Qiskit, Cirq, etc.). &$\square$ \\
& The version of the software framework and any dependencies &$\square$ \\
& Random seeds used in case of simulation-based experiments &$\square$ \\
& Measurement settings and readout error mitigation techniques. &$\square$ \\ \hline
\multirow{2}{*}{Analysis} & Statistical modelling of the distributions of outputs &$\square$ \\
& Quantum error correction or mitigation techniques employed &$\square$ \\ \hline
Treats to  &Analysis of how the results might vary with different quantum hardware &$\square$ \\
Validity& Description for how the experiment might be modified or extended to other quantum computing platforms &$\square$ \\ \hline
\end{tabular}
}
\caption{Checklist for Quantum Computing Experiments reporting}
\label{table:checklist}
\end{table*}

\section*{Appendix B: A checklist for the laboratory packages of quantum software experiments}

\begin{table}[h]
\centering
\resizebox{0.6\textwidth}{!}{
\begin{tabular}{|p{2cm}|p{10.5cm}|c|}
\hline
    Article & Scientific article & $\square$  \\
    & Description of the experimental design & $\square$ \\
    & Description of the experiment context & $\square$ \\
    & Authors & $\square$ \\ \hline

    Experimental & Instructions for the experimental material & $\square$ \\
    Material & Conceptual Models and/or representative diagrams & $\square$ \\ 
    & Materials needed to conduct the experiment (source code, quantum computer configuration or provider, features, etc.) & $\square$ \\ \hline

    Datasets & Pre-experimental data & $\square$ \\
    & Data processing & $\square$ \\
    & Data resulting from the experiment & $\square$ \\ \hline

    System & Selection and definition of hardware technical requirements (topology, connection, number of qubits, queue, cost, etc.) & $\square$ \\
    & Conditions and restrictions of the quantum computer and/or provider (maximum number of qubits, availability, decoherence, etc.) & $\square$ \\
    & Description of Classic Software & $\square$ \\
    & Description of Quantum Software (gates/annealing, algorithm, number of shots) & $\square$ \\ \hline
    
    Scripts & Description of the followed script (including decisions made)  & $\square$ \\
    & Script for analyzing the obtained results & $\square$ \\ 
    & Log file of the obtained results (CSV, images, etc.) & $\square$ \\ \hline
    
    Dictionary & Description of the artifacts & $\square$ \\
    & Artifacts inventory (Bitbucket, GitHub, etc.) & $\square$ \\
    & Instructions for downloading artifacts (classic and/or quantum) & $\square$ \\
    & Instruction guide for executing artifacts (classic and/or quantum) & $\square$ \\
    & Artifact license & $\square$ \\
    & Virtual environments (Docker, Kubernetes, etc.) & $\square$ \\
    & Definition of the development environment & $\square$ \\
    & Guidelines for configuring and accessing the development environment & $\square$ \\
    & Verification of access and functionality & $\square$ \\
    & Decisions made during the experiment & $\square$ \\
    & DOI & $\square$ \\
    & License or distribution right of the Laboratory Package & $\square$ \\ \hline
\end{tabular}
}
\caption{Checklist for Quantum Computing Laboratory Packages}
\label{table:checklistLP}
\end{table}

\end{document}

%% file: intro.tex
\section{Introduction}
\label{sec:introduction}

Quantum computing is an emergent paradigm that holds the potential to address challenges that conventional computing, in its current form, struggles to resolve within a practical timeframe \cite{Harrow2017}. Quantum Computing introduces a paradigm shift by harnessing the principles of quantum mechanics, offering the prospect of tackling computational problems that were previously deemed insurmountable within traditional computing frameworks.
Today, there are already advances in the field of Quantum Software Engineering and there are technological proposals and experiments in this new area of Software Engineering \cite{Serrano2022}. However, as in traditional Empirical Software Engineering, conducting an experiment in Quantum Software Engineering faces conceptual and material barriers that must be overcome 
\cite{Piattini2020}.

To perform such experiments, researchers must manage a number of concepts intrinsic to quantum computing. Materially, these  experiments require acces to a quantum computer, which usually involves using such hardware owned by other organizations as a service on a pay-per-use basis, or access to a quantum simulator
. Due to these distinct requirements, experimentation in quantum computing diverges from classical software engineering. This divergence necessitates the development of novel techniques, mechanisms, and methodologies, or the adaptation of existing approaches to align with the unique characteristics of this new computing paradigm.

Experts in the field of Quantum Software Engineering believe that empirical software engineering methods and techniques need to be revised to meet industry expectations in relation to quantum computing, making the development of this new area in a priority \cite{Murillo2024}.

A \emph{Quantum Software Experiment (QSE)} is characterized by the integration of quantum software within the experimental framework, where its role can vary significantly. This role may range from being a critical component essential to achieving the experiment's objectives and influencing its variables, to a more nuanced function.
Accurate and thorough reporting of the experiment’s details, context, execution, and results is crucial for ensuring its validity and replicability. Complementing this, a Laboratory Package (LP) is a collection of artifacts systematically employed throughout the entire lifecycle of an experiment, encompassing the planning, execution, and analysis stages \cite{Shull2002}.  

For QSEs, the reporting and replicability problem is even more challenging. The current state of quantum computing entails that software is often developed through the specification of circuits based on logic gates, and not all quantum software can be executed on any of the existing quantum computers due to the significant technological diversity \cite{alvarado2023}. 
This variability has fuelled numerous proposals aimed at simplifying the implementation and deployment of quantum software. \cite{Romero-Alvarez2023,alvarado2023}, but they also affect the requirements for reporting QSEs, threatening their applicability \cite{Hirayama2022}. 

In this work, we aim to facilitate the replication of QSEs, since despite the unique characteristics and complexities associated with quantum software that demands meticulous attention in experimental reporting, there remains a noticeable absence of tailored guidelines.  This paper tries to fill this gap, by developing specific reporting guidelines based on the previous generic proposals created for empirical software engineering, and addressing the nuanced requirements of quantum computing research. 
To fully satisfy this objective, we not only propose a set of guidelines for the reporting of quantum software experiments, but we also outline a laboratory package structure for experiments in this area.

Thus, this paper's contributions are twofold: Firstly, it offers an extension of the established reporting guidelines for software engineering experiments stated in \cite{jedlitschka2008reporting}, specifically tailored for quantum software engineering. Secondly, it proposes a comprehensive LP structure, encompassing all necessary components for conducting and replicating experiments in quantum computing. This package aims to facilitate rigorous and reproducible research within this emerging field.


The rest of the paper is structured as follows: Section \ref{sec:background} provides an overview of essential quantum computing concepts and relevant previous studies. Sections \ref{Reporting}  and \ref{LaboratoryPackage}  respectively present our proposed experiment reporting guidelines and proposed integrated LP structure for the quantum software engineering domain. Section \ref{sec:validation} details the validation carried out for our proposal. Finally, Section \ref{sec:conclusion} presents a discourse on the implications and conclusions drawn from our research.